\documentclass[a4paper, twocolumn]{article}
\usepackage[english]{babel}
\usepackage[utf8x]{inputenc}
\usepackage[T1]{fontenc}

\usepackage[a4paper,top=3cm,bottom=2cm,left=3cm,right=3cm,marginparwidth=1.75cm]{geometry}

\usepackage[labelfont=bf]{caption}
\DeclareCaptionLabelSeparator{pipe}{ | }
\usepackage{amsmath}
\usepackage{graphicx}
\graphicspath{ {images/} }
\usepackage{subfig}
\usepackage{caption}
\usepackage{bm}
\usepackage[colorinlistoftodos]{todonotes}
\usepackage[colorlinks=true, allcolors=blue]{hyperref}
\usepackage{color}

\usepackage[superscript]{cite}
\makeatletter
\renewcommand\@biblabel[1]{#1.}
\makeatother

\title{Multiparameter optimisation of a magneto-optical trap using deep learning}
\author{A. D. Tranter \and H. J. Slatyer \and M. R. Hush \and A. C. Leung \and J. L. Everett \and K. V. Paul \and P. Vernaz-Gris \and P. K. Lam \and B. C. Buchler \and G. T. Campbell}

\begin{document}
\maketitle

\textbf{
Many important physical processes have dynamics that are too complex to completely model analytically. Optimisation of such processes often relies on intuition, trial-and-error, or the construction of empirical models. Machine learning based on artificial neural networks has emerged as an efficient means to develop empirical models of complex systems. We implement a deep artificial neural network to optimise the magneto-optic cooling and trapping of neutral atomic ensembles. Cold atomic ensembles have become commonplace in laboratories around the world, however, many-body interactions give rise to complex dynamics that preclude precise analytic optimisation of the cooling and trapping process. The solution identified by machine learning is radically different to the smoothly varying adiabatic solutions currently used. Despite this, the solutions vastly outperform best known solutions producing higher optical densities. This may provide a pathway to a new understanding of the dynamics of the cooling and trapping processes in cold atomic ensembles.}

The interaction of light and atoms has long been a valuable test-bed for the foundations of quantum mechanics. The laser cooling of atoms \cite{Migdall:1985bs,Phillips:1998hw} was a turning point that enabled a range of new and exciting developments in atom-light coupling techniques, shedding the complications inherent in the motion of free atoms. Cold atomic ensembles underpin many important advances such as the generation of Bose-Einstein condensates \cite{Cornell:2002ir,Ketterle:2002jm}, cold atom based precision metrology \cite{Kitching:2011kb}, a new generation of optical atomic clocks \cite{Gobel:2015cw} and quantum information processing \cite{briegel1998quantum,Dowling:2003ei,lvovsky2009optical,sangouard2011quantum}.

\begin{figure*}[!ht]
\centering
\includegraphics[width=\textwidth]{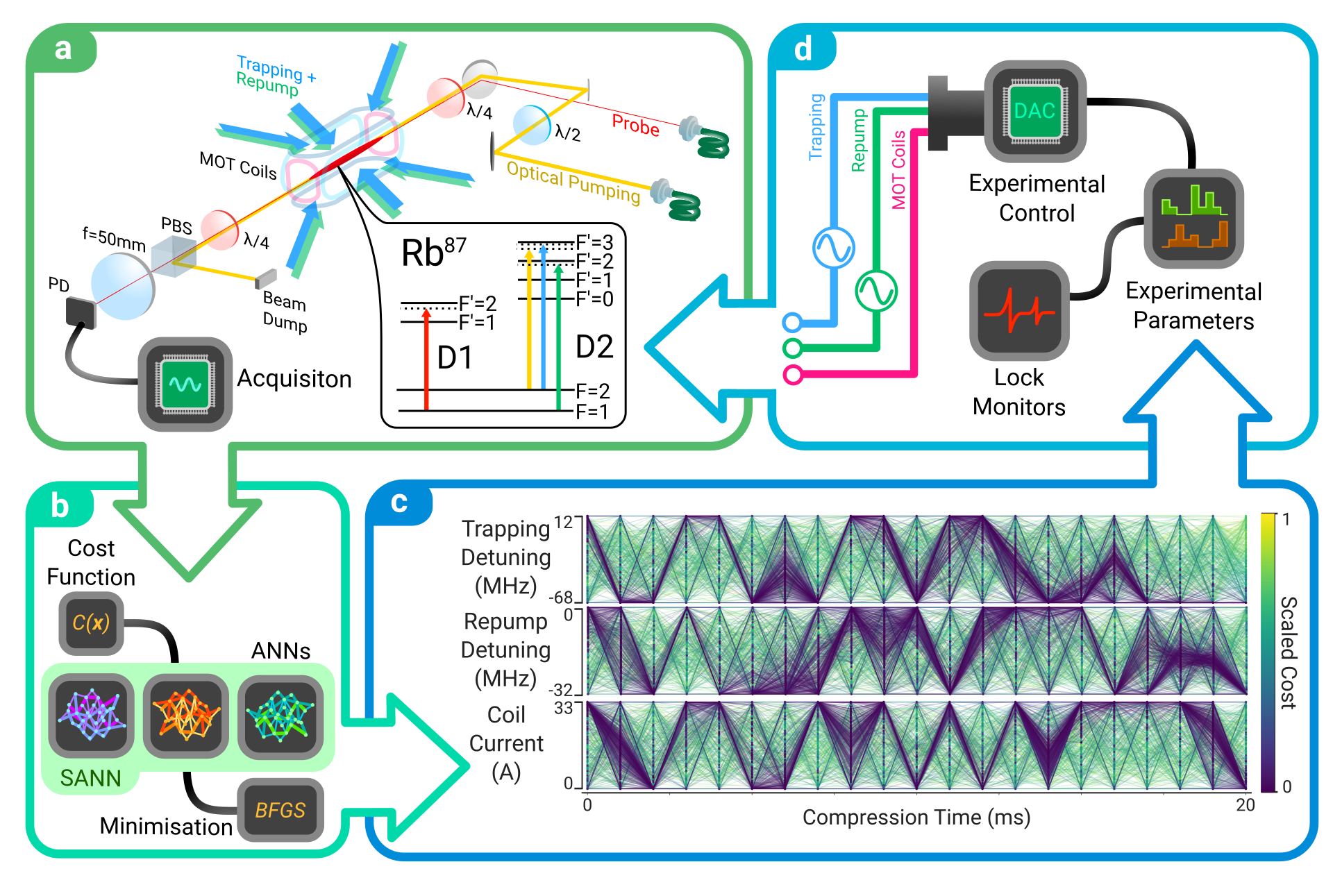}
\caption{\textbf{Online optimisation of optical depth.} \textbf{a}, Initially a MOT captures thermal Rb\textsuperscript{87} atoms via laser cooling. The ensemble is then transiently compressed using a set of 21 time bins for trapping frequency, repump frequency and magnetic field strength. The off resonant OD is measured from the transmission of a probe field incident on a photo detector. This value is passed to the SANN \textbf{b}, where a cost function is calculated for the current set of parameters. Each ANN that comprises the SANN is trained using this and the previous training data. Each ANN generates a parameter set by minimising the predicted cost landscape using the Broyden–Fletcher–Goldfarb–Shanno (BFGS) algorithm. This generates a new set of parameters \textbf{c}, which facilitates exploration of the landscape. Here a scaled cost of 1 denotes a failure to trap atoms with a given parameter configuration and 0 corresponds to the best known parameter set. Each predicted parameter set is sequentially passed to \textbf{d}, the experimental control systems which monitor the lock state of the experiment and convert the parameter set to physical values. This loop continues until either a minimisation condition or maximum run number is reached.\label{fig:exp_setup}}
\end{figure*}

In general, the efficacy of cold atomic ensembles is improved by increasing the number of atoms and reducing the temperature as this will increase signal-to-noise ratio of any measurement of the atoms.  
In the particular example of quantum information processing, the key metric is the resonant optical depth (OD) of the ensemble. The larger the OD, the stronger the atom-light coupling which is essential for maintaining coherence while mapping into and out of the atoms. The most prevalent cold atomic system is that of the magneto-optical trap (MOT) (see Fig. \ref{fig:exp_setup}a) in which thermal atoms are collected into the trap from the surrounding warm vapour. OD is highly dependent on atomic species and trap geometry, however there exists a number of strategies to improve this characteristic, such as transient compression stages within an experimental run \cite{petrich1994behavior,depue2000transient}, polarisation gradient cooling \cite{dalibard1989laser} and temporal/spatial dark spots \cite{depue2000transient}. Despite the extensive amount of work done on laser-cooled atomic systems \cite{grynberg2001cold}, it remains a challenging endeavour to construct a quantitative description that captures the complete atomic dynamics. This is mostly owing to the fact that these systems generally present computationally intractable dynamics in 3-dimensions, involving many body interactions, polarisation gradients and complex scattering processes \cite{townsend1995phase, hanley2017quantitative}. Furthermore, analytical models fail to account for experimental imperfections that may perturb the system. As such most strategies to improve optical depth are in general limited to intuition regarding adiabatic and monotonic approaches. However there have been indications that solutions outside of this space may lead to more efficient collection of atomic ensembles \cite{torrontegui2013shortcuts}. Recently it was also demonstrated that it is possible that BECs may be distilled from cold ensembles without evaporative cooling techniques by using a specialised compression sequence \cite{hu2017creation}.

Machine learning techniques, in particular those based on deep neural networks (``deep learning''), have shown great promise for solving complex problems beyond human performance \cite{silver2016mastering,mnih2015human}. 
In the current work we seek to optimise the OD of a cold atomic ensemble by tuning the compression sequence using an algorithm based on deep learning. Such algorithms have not previously been applied to quantum systems. Our approach involves a feedback-like online procedure in which the algorithm takes control of 63 independent piecewise experimental parameters and automatically adjusts them to optimise the system. Often human intuition regarding a given system can be difficult to achieve and in some cases misleading. For such systems, optimisation via an algorithmic process designed only to minimise a cost function can identify solutions that are highly non-intuitive and yet outperform traditional solutions.

Machine learning techniques have been used to optimize the control of quantum experiments and theoretical protocols \cite{palittapongarnpim2017learning,palittapongarnpim2016controlling,li2017rapid,august2017using,ju2017designing,mavadia2017prediction}. However in order to probe the complex dynamics of the MOT we require an automated online optimisation platform. Gaussian process (GP) models have been used to perform online optimisation in an ultra-cold atom experiment \cite{wigley2016fast}, however, the time required to fit a GP scales with the cube of the number of experimental runs \cite{barber2012bayesian}. This would quickly exceed the time for an individual experimental acquisition for our larger number of parameters. Using an artificial neural network (ANN) allows us to significantly increase the number of experimentally controlled parameters as the training time for an ANN scales linearly with number of performed trials.

Evolutionary algorithms have been applied to the optimisation of cold-atomic and quantum systems \cite{amstrup1995genetic,geisel2013evolutionary,beil2010light,tsubouchi2008rovibrational,warren1993coherent} and have used ANNs as "surrogates" to accelerate the convergence \cite{jin2011surrogate,jin2005comprehensive,won2005framework} in classical control problems, such as microwave engineering \cite{rayas2004based} and airfoil design \cite{hacioglu2007fast}. In all these cases, the evolutionary algorithm is put in charge of picking the next points in the experiment and balancing the exploration vs. exploitation trade-off \cite{kaelbling1996reinforcement}. 
Our approach is to  use the probabilistic predictions of a stochastic artificial neural network (SANN) together with the Thompson sampling technique \cite{thompson1933on} to balance this trade-off. This effectively puts the SANN in control of the experiment, instead of acting as a ``surrogate'' used only to quickly evaluate an approximation of the cost function. To our knowledge, this work presents the only ANN that directly automatically optimizes an experiment, with the largest number of parameters.

In the present work we use densely connected multilayer perceptrons as models of the MOT response to a given set of parameters. Our topology consists of a 5-hidden layer network with 64 neurons each, which can be trained in under one second on standard hardware (Intel i7-920 2.67GHz). Our choice of activation function is the Gaussian error linear unit, which yields fast training and smooth landscapes \cite{hendrycks2016bridging}. The ANNs are trained using the Adam algorithm, which adaptively sets the step size. The early stopping technique is used to avoid over-fitting. To facilitate exploration of the landscape, the algorithm must also choose the next set of parameters to explore. The algorithm achieves this using the Thompson sampling technique, whereby a deterministic model is sampled from the SANN and the best parameters predicted by that model are chosen. To realise the probabilistic predictions of our SANN we implement bagging \cite{breiman1996bagging} with an ensemble of three ANNs. As we have relatively few data points, each network is trained on the full dataset and we rely on the independent initialisations and inherent randomness in the training procedure to provide the variety in our models. Once a model has been sampled it is then probed for minima using the L-BFGS-B algorithm \cite{Byrd:2006iv}.

\begin{figure*}[ht]
\centering
\includegraphics[width=\textwidth]{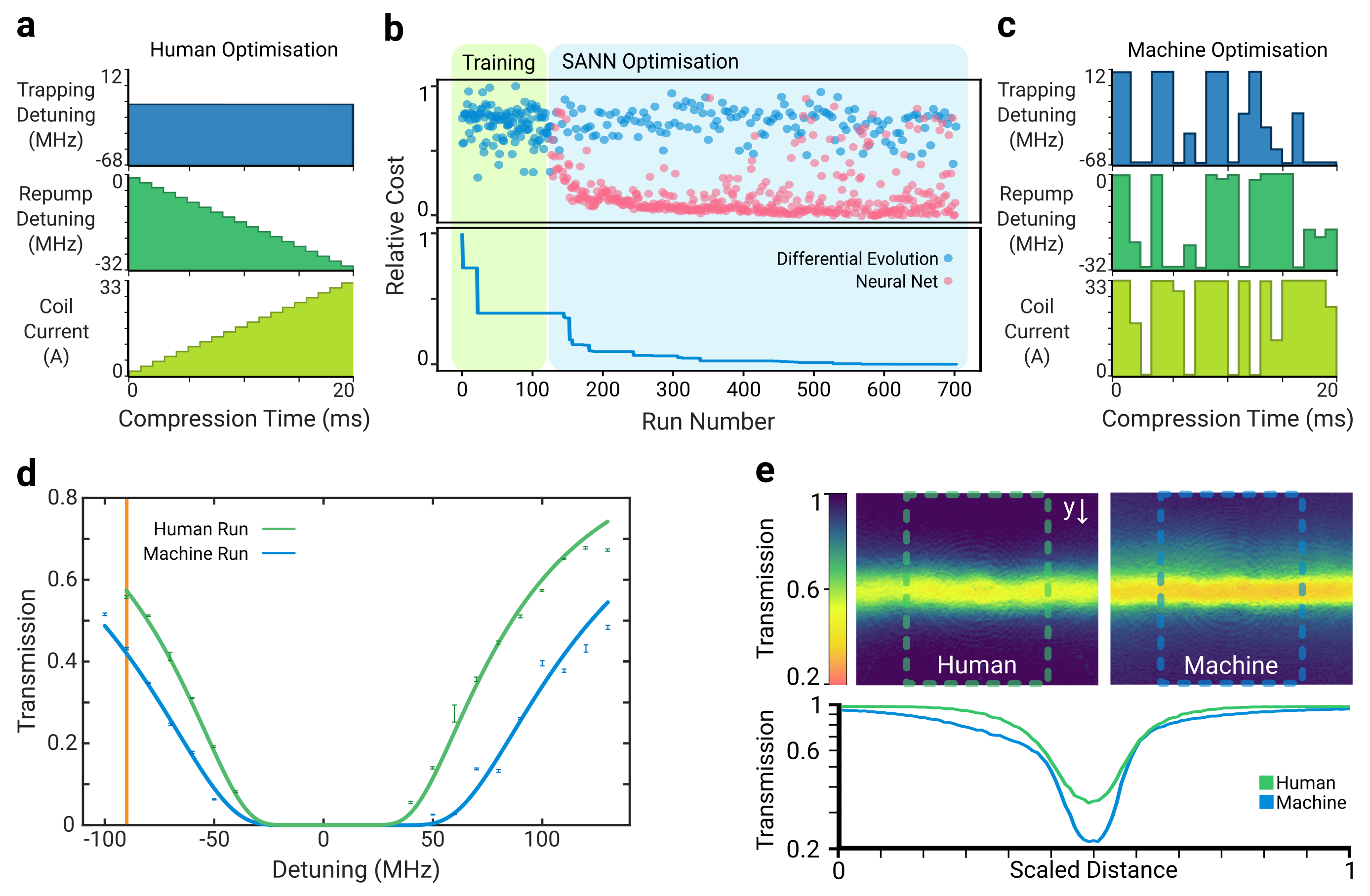}
\caption{\textbf{Experimental results using the SANN optimisation.} \textbf{a}, Human optimised compression stage using monotonic ramps for the magnetic fields and repump frequency (temporal dark SPOT). \textbf{b}, Convergence of the SANN on an optimal solution after 126 training runs. The pink points are predictions generated by the ANNs while the blue points are generated by the differential evolution algorithm geared towards exploration. \textbf{c}, Solution generated by the SANN for 63 discrete parameters that maximises off resonant OD by minimising the transmitted probe field. \textbf{d}, OD measurements for the human and SANN optimised ensembles for a wide range of detunings corresponding to $530\pm8$ and $970\pm20$ respectively. The vertical line indicates the detuning used for optimisation. \textbf{e}, Absorption images for the human and SANN optimised atomic ensembles. The lower log-plot shows a cross-section of the spatial distribution of the atoms which is directly influenced by each compression sequence. The boxes on the images indicate the integration windows used for the cross-sections. \label{fig:results}}
\end{figure*}

Initially the SANN is trained using a differential evolution (DE) algorithm biased towards exploration for 2N points, where N is the number of parameters. After this each of the three ANNs generates a prediction to be explored, with every fourth point predicted by the DE algorithm to ensure the ANNs continually receive unbiased data. These new data points are added to the training data and the SANN is trained again on the new data set. This cycle is repeated until the minimum cost is determined.

We apply our implementation of a SANN to experimentally optimising the optical depth (OD) of a cold atomic ensemble. Our system comprises of a Rb\textsuperscript{87} elongated MOT as shown in Fig. \ref{fig:exp_setup} used for quantum memory experiments. The elongated shape provides the high OD required for such quantum memory experiments and is achieved via 2D elongated trapping coils \cite{Sparkes:2013jw,cho2016highly} and additional capping coils for axial confinement. A detuned probe beam is sent through the atomic ensemble as a quick proxy for measuring OD (see Methods). This measurement is fed back into the main optimisation loop so that the next set of parameters may be selected.
We initially optimised the compression sequence following the conventional monotonic approaches used in other works \cite{sparkes2013gradient,depue2000transient,ketterle1993high} (See Fig. \ref{fig:results}a). As this period of the experiment is crucial to the final optical density of the ensemble and spans a relatively short time compared to other periods within an experimental run, this sequence was chosen as the platform for online optimisation. We divide our compression sequence into 21 sequential time bins with a period of 1ms each. We provide the SANN with arbitrary experimental control of the trapping field detuning, repump field detuning and magnetic field strength during each of these time bins giving a 63 parameter optimisation. These parameters are expected to have the largest effect on the transient compression of the ensemble as each parameter contributes directly to physical characteristics of the ensemble and trap such as scattering rate, cooling rate, velocity capture range and density. Furthermore, the SANN also has control of the detuning during the polarisation gradient cooling stage by way of setting the final value of the trapping detuning at the end of the compression sequence. In general we are limited by the experimental duty cycle and not computational power with 63 parameters providing a good trade-off between ramp granularity and the parameter landscape size.

The figure of merit or cost function for this optimisation is monitored via a photodetector which measures the incident probe field after absorption by the atomic ensemble. Large focusing optics are employed to eliminate the effects of lensing which may distort the measured OD profile, as this could allow a "gaming" of the cost function. As this optimisation is concerned purely with OD we construct a simple cost function

\begin{equation}
C(\bm{X}) = \frac{1}{P} \int ^{t_{f}} _{t_{i}} p(t) \, dt  
\end{equation}

where $C(\bm{X})$ is the cost for a given set of parameters $\bm{X}$, $P$ is a scaling factor derived from a reference signal to correct fluctuations in laser power and $p(t)$ is the measured photodector response which is integrated across the pulse window. In this way a larger(smaller) cost is attributed to higher(lower) transmission of the incident probe light and thus lower(higher) OD. While this is not a strict measurement of OD, this inferred quantity affords a quick approximation which allows the optimisation speed to remain close to that of the repetition rate of the experiment.

We wish to optimise the transient optical depth of the ensemble while maintaining a high repetition rate. To this end, we measure the cost function with the experiment repeating the compression sequence at a rate of 2 Hz. At this repetition rate, atoms lost from the trap during one compression sequence will negatively affect the optical depth for subsequent compressions. The SANN must therefore reach a solution that not only provides a large transient compression, but also avoids atom loss. When applying each new set of parameters to test, the ensemble is released and we wait until the trap refills to a steady-state to ensure the next measurement is not biased by the previous parameter set. This wait time limits the rate at which we can test parameter sets to roughly once per ten seconds.

An initial 126 training runs were collected which formed the training set used to concurrently train the ANNs. Following this a further 577 runs were recorded consisting of predictions made by the SANN. Convergence on the optimal solution occurred after approximately 520 runs as show in Fig. \ref{fig:results}b, with the optimal solution given in Fig. \ref{fig:results}c.  It can be seen from Fig. \ref{fig:exp_setup}c that our parameter space is not monotonic or flat, with a scaled cost of 1 corresponding to the absence of an atomic ensemble due to a failure to trap atoms with the parameter configuration.

We find that each of the ANNs converge to a distinct set of parameters corresponding to the optimal measured cost, while there exists many sub-optimal parameter sets which result in a less optically dense ensemble. Subsequent optimisation runs revealed that our parameter landscape contained many local minima which were within a few percent of the best presented solution. While different optimisations converged on these local minima we find that the relatively effectiveness of each solution is constant irrespective of daily experimental drift. 
From inspection of the human and SANN generated solutions shown in Fig. \ref{fig:results}a,c, it is immediately obvious that the SANN solution is radically different to conventional techniques regarding temporal dark SPOTs \cite{ketterle1993high} and transient compression schemes \cite{depue2000transient,sparkes2013gradient} that seek to increase atomic density. Instead we demonstrate a solution that exhibits structures that swing between the boundary values of our experimental setup and show little regard for continuity or monotonicity. While these piecewise ramps seem to defy physical intuition, we find that they invariably outperform the best human solutions. The complex structures of the solutions hint at dynamics that are not well understood, although we speculate that they may be related to release-and-capture dynamics that have been observed in optical lattices \cite{depue1999unity}.

As the measured cost is not a true measurement of OD we also sought to characterise the OD by mapping the absorption curve as a function of probe detuning. As shown in Fig. \ref{fig:results}d we find good agreement between experimental data and the theoretical absorption given by:

\begin{equation}
I_t/I_0 = e^{OD\frac{\gamma^2 / 4}{\Delta^2 + \gamma^2/4}}
\end{equation}

where $I_t/I_0$ is the normalised transmitted intensity, OD is optical depth, $\gamma$ is the excited state decay rate and $\Delta$ is the probe detuning. We measure the OD of the best human optimised solution to be $535 \pm 8$ while the ANN optimised solution gives a measured OD of $970 \pm 20$, supporting the ANN solution as the optimal solution. The OD achieved by the human solution is on the order of our previously reported results used for high efficiency memory experiments \cite{cho2016highly} and represents the best achievable OD with our system using current methods. The ANN solution however affords an OD increase of $(81\pm3)\%$ by being agnostic to these techniques. Furthermore we find that absorption imaging through the side of the cloud shows a clear physical distinction between the solutions. We image using an expanded beam on the repump transition 9 MHz red detuned. As shown in Fig. \ref{fig:results}e the ANN optimised solution has a higher density of atoms around the center of the probe axis as well as increased homogeneity along the longitudinal axis. We also note that the atomic distribution is modified with a leading tail corresponding to a halo of atoms collected around the top of the atomic cloud.

\paragraph{Conclusions}
In summary, we have used a stochastic artificial neural network (SANN) to build a black-box empirical model of the cooling and trapping of neutral atoms, a process that is often difficult to model precisely using theoretical methods. Our solutions improve on those born from theoretical considerations regarding adiabatic compression. 
While we present a specific solution, we do not tailor the SANN to the experiment, which leaves open the possibility to apply these methods to a wide range of experimental setups and problems. We find that the solutions are robust to daily fluctuations within the experiment and retain their relative efficacy. In general we are limited by the experimental duty cycle, so it is feasible that the SANN can be applied to experiments with larger parameter landscapes provided that they also support a higher duty cycle. We believe this would be well suited to applications with high dimensional structures such as imaging.

\small
\bibliographystyle{naturemag.bst}
\bibliography{main}

\begin{thebibliography}{10}
\expandafter\ifx\csname url\endcsname\relax
  \def\url#1{\texttt{#1}}\fi
\expandafter\ifx\csname urlprefix\endcsname\relax\def\urlprefix{URL }\fi
\providecommand{\bibinfo}[2]{#2}
\providecommand{\eprint}[2][]{\url{#2}}

\bibitem{Migdall:1985bs}
\bibinfo{author}{Migdall, A.~L.}, \bibinfo{author}{Prodan, J.~V.},
  \bibinfo{author}{Phillips, W.~D.}, \bibinfo{author}{Bergeman, T.~H.} \&
  \bibinfo{author}{Metcalf, H.~J.}
\newblock \bibinfo{title}{{First Observation of Magnetically Trapped Neutral
  Atoms}}.
\newblock \emph{\bibinfo{journal}{Physical Review Letters}}
  \textbf{\bibinfo{volume}{54}}, \bibinfo{pages}{2596--2599}
  (\bibinfo{year}{1985}).

\bibitem{Phillips:1998hw}
\bibinfo{author}{Phillips, W.~D.}
\newblock \bibinfo{title}{{Nobel Lecture: Laser cooling and trapping of neutral
  atoms}}.
\newblock \emph{\bibinfo{journal}{Reviews of Modern Physics}}
  \textbf{\bibinfo{volume}{70}}, \bibinfo{pages}{721--741}
  (\bibinfo{year}{1998}).

\bibitem{Cornell:2002ir}
\bibinfo{author}{Cornell, E.~A.} \& \bibinfo{author}{Wieman, C.~E.}
\newblock \bibinfo{title}{{Nobel Lecture: Bose-Einstein condensation in a
  dilute gas, the first 70 years and some recent experiments}}.
\newblock \emph{\bibinfo{journal}{Reviews of Modern Physics}}
  \textbf{\bibinfo{volume}{74}}, \bibinfo{pages}{875--893}
  (\bibinfo{year}{2002}).

\bibitem{Ketterle:2002jm}
\bibinfo{author}{Ketterle, W.}
\newblock \bibinfo{title}{{Nobel lecture: When atoms behave as waves:
  Bose-Einstein condensation and the atom laser}}.
\newblock \emph{\bibinfo{journal}{Reviews of Modern Physics}}
  \textbf{\bibinfo{volume}{74}}, \bibinfo{pages}{1131} (\bibinfo{year}{2002}).

\bibitem{Kitching:2011kb}
\bibinfo{author}{Kitching, J.}, \bibinfo{author}{Knappe, S.} \&
  \bibinfo{author}{Donley, E.~A.}
\newblock \bibinfo{title}{{Atomic Sensors {\textendash} A Review}}.
\newblock \emph{\bibinfo{journal}{IEEE Sensors Journal}}
  \textbf{\bibinfo{volume}{11}}, \bibinfo{pages}{1749--1758}
  (\bibinfo{year}{2011}).

\bibitem{Gobel:2015cw}
\bibinfo{author}{G{\"o}bel, E.~O.} \& \bibinfo{author}{Siegner, U.}
\newblock \emph{\bibinfo{title}{{Laser Cooling, Atomic Clocks, and the
  Second}}}, vol.~\bibinfo{volume}{54} of \emph{\bibinfo{series}{Gobel/Quantum
  Metrology: Foundation of Units and Measurements}}
  (\bibinfo{publisher}{Wiley-VCH Verlag GmbH {\&} Co. KGaA},
  \bibinfo{address}{Weinheim, Germany}, \bibinfo{year}{2015}).

\bibitem{briegel1998quantum}
\bibinfo{author}{Briegel, H.-J.}, \bibinfo{author}{Dür, W.},
  \bibinfo{author}{Cirac, J.~I.} \& \bibinfo{author}{Zoller, P.}
\newblock \bibinfo{title}{{Quantum repeaters: the role of imperfect local
  operations in quantum communication}}.
\newblock \emph{\bibinfo{journal}{Physical Review Letters}}
  \textbf{\bibinfo{volume}{81}}, \bibinfo{pages}{5932} (\bibinfo{year}{1998}).

\bibitem{Dowling:2003ei}
\bibinfo{author}{Dowling, J.~P.} \& \bibinfo{author}{Milburn, G.~J.}
\newblock \bibinfo{title}{{Quantum technology: the second quantum revolution}}.
\newblock \emph{\bibinfo{journal}{Philosophical Transactions of the Royal
  Society of London A: Mathematical, Physical and Engineering Sciences}}
  \textbf{\bibinfo{volume}{361}}, \bibinfo{pages}{1655--1674}
  (\bibinfo{year}{2003}).

\bibitem{lvovsky2009optical}
\bibinfo{author}{Lvovsky, A.~I.}, \bibinfo{author}{Sanders, B.~C.} \&
  \bibinfo{author}{Tittel, W.}
\newblock \bibinfo{title}{{Optical quantum memory}}.
\newblock \emph{\bibinfo{journal}{Nature photonics}}
  \textbf{\bibinfo{volume}{3}}, \bibinfo{pages}{706} (\bibinfo{year}{2009}).

\bibitem{sangouard2011quantum}
\bibinfo{author}{Sangouard, N.}, \bibinfo{author}{Simon, C.},
  \bibinfo{author}{{De Riedmatten}, H.} \& \bibinfo{author}{Gisin, N.}
\newblock \bibinfo{title}{{Quantum repeaters based on atomic ensembles and
  linear optics}}.
\newblock \emph{\bibinfo{journal}{Reviews of Modern Physics}}
  \textbf{\bibinfo{volume}{83}}, \bibinfo{pages}{33} (\bibinfo{year}{2011}).

\bibitem{petrich1994behavior}
\bibinfo{author}{Petrich, W.}, \bibinfo{author}{Anderson, M.~H.},
  \bibinfo{author}{Ensher, J.~R.} \& \bibinfo{author}{Cornell, E.~A.}
\newblock \bibinfo{title}{Behavior of atoms in a compressed magneto-optical
  trap}.
\newblock \emph{\bibinfo{journal}{JOSA B}} \textbf{\bibinfo{volume}{11}},
  \bibinfo{pages}{1332--1335} (\bibinfo{year}{1994}).

\bibitem{depue2000transient}
\bibinfo{author}{DePue, M.~T.}, \bibinfo{author}{Winoto, S.~L.},
  \bibinfo{author}{Han, D.} \& \bibinfo{author}{Weiss, D.~S.}
\newblock \bibinfo{title}{{Transient compression of a MOT and high intensity
  fluorescent imaging of optically thick clouds of atoms}}.
\newblock \emph{\bibinfo{journal}{Optics communications}}
  \textbf{\bibinfo{volume}{180}}, \bibinfo{pages}{73–79}
  (\bibinfo{year}{2000}).

\bibitem{dalibard1989laser}
\bibinfo{author}{Dalibard, J.} \& \bibinfo{author}{Cohen-Tannoudji, C.}
\newblock \bibinfo{title}{{Laser cooling below the Doppler limit by
  polarization gradients: simple theoretical models}}.
\newblock \emph{\bibinfo{journal}{JOSA B}} \textbf{\bibinfo{volume}{6}},
  \bibinfo{pages}{2023–2045} (\bibinfo{year}{1989}).

\bibitem{grynberg2001cold}
\bibinfo{author}{Grynberg, G.} \& \bibinfo{author}{Robilliard, C.}
\newblock \bibinfo{title}{Cold atoms in dissipative optical lattices}.
\newblock \emph{\bibinfo{journal}{Physics Reports}}
  \textbf{\bibinfo{volume}{355}}, \bibinfo{pages}{335--451}
  (\bibinfo{year}{2001}).

\bibitem{townsend1995phase}
\bibinfo{author}{Townsend, C.} \emph{et~al.}
\newblock \bibinfo{title}{{Phase-space density in the magneto-optical trap}}.
\newblock \emph{\bibinfo{journal}{Physical Review A}}
  \textbf{\bibinfo{volume}{52}}, \bibinfo{pages}{1423} (\bibinfo{year}{1995}).

\bibitem{hanley2017quantitative}
\bibinfo{author}{Hanley, R.~K.} \emph{et~al.}
\newblock \bibinfo{title}{Quantitative simulation of a magneto-optical trap
  operating near the photon recoil limit}.
\newblock \emph{\bibinfo{journal}{arXiv preprint arXiv:1706.04807}}
  (\bibinfo{year}{2017}).

\bibitem{torrontegui2013shortcuts}
\bibinfo{author}{Torrontegui, E.} \emph{et~al.}
\newblock \bibinfo{title}{{Shortcuts to adiabaticity}}.
\newblock \emph{\bibinfo{journal}{Adv. At. Mol. Opt. Phys}}
  \textbf{\bibinfo{volume}{62}}, \bibinfo{pages}{117–169}
  (\bibinfo{year}{2013}).

\bibitem{hu2017creation}
\bibinfo{author}{Hu, J.} \emph{et~al.}
\newblock \bibinfo{title}{Creation of a bose-condensed gas of 87rb by laser
  cooling}.
\newblock \emph{\bibinfo{journal}{Science}} \textbf{\bibinfo{volume}{358}},
  \bibinfo{pages}{1078--1080} (\bibinfo{year}{2017}).

\bibitem{silver2016mastering}
\bibinfo{author}{Silver, D.} \emph{et~al.}
\newblock \bibinfo{title}{Mastering the game of go with deep neural networks
  and tree search}.
\newblock \emph{\bibinfo{journal}{nature}} \textbf{\bibinfo{volume}{529}},
  \bibinfo{pages}{484--489} (\bibinfo{year}{2016}).

\bibitem{mnih2015human}
\bibinfo{author}{Mnih, V.} \emph{et~al.}
\newblock \bibinfo{title}{Human-level control through deep reinforcement
  learning}.
\newblock \emph{\bibinfo{journal}{Nature}} \textbf{\bibinfo{volume}{518}},
  \bibinfo{pages}{529} (\bibinfo{year}{2015}).

\bibitem{palittapongarnpim2017learning}
\bibinfo{author}{Palittapongarnpim, P.}, \bibinfo{author}{Wittek, P.},
  \bibinfo{author}{Zahedinejad, E.}, \bibinfo{author}{Vedaie, S.} \&
  \bibinfo{author}{Sanders, B.~C.}
\newblock \bibinfo{title}{Learning in quantum control: high-dimensional global
  optimization for noisy quantum dynamics}.
\newblock \emph{\bibinfo{journal}{Neurocomputing}}
  \textbf{\bibinfo{volume}{268}}, \bibinfo{pages}{116--126}
  (\bibinfo{year}{2017}).

\bibitem{palittapongarnpim2016controlling}
\bibinfo{author}{Palittapongarnpim, P.}, \bibinfo{author}{Wittek, P.} \&
  \bibinfo{author}{Sanders, B.~C.}
\newblock \bibinfo{title}{Controlling adaptive quantum phase estimation with
  scalable reinforcement learning}.
\newblock In \emph{\bibinfo{booktitle}{24th European Symposium on Artificial
  Neural Networks, Bruges, April 27--29, 2016}}, \bibinfo{pages}{327--332}
  (\bibinfo{year}{2016}).

\bibitem{li2017rapid}
\bibinfo{author}{Li, C.} \emph{et~al.}
\newblock \bibinfo{title}{Rapid bayesian optimisation for synthesis of short
  polymer fiber materials}.
\newblock \emph{\bibinfo{journal}{Scientific reports}}
  \textbf{\bibinfo{volume}{7}}, \bibinfo{pages}{5683} (\bibinfo{year}{2017}).

\bibitem{august2017using}
\bibinfo{author}{August, M.} \& \bibinfo{author}{Ni, X.}
\newblock \bibinfo{title}{Using recurrent neural networks to optimize dynamical
  decoupling for quantum memory}.
\newblock \emph{\bibinfo{journal}{Physical Review A}}
  \textbf{\bibinfo{volume}{95}}, \bibinfo{pages}{012335}
  (\bibinfo{year}{2017}).

\bibitem{ju2017designing}
\bibinfo{author}{Ju, S.} \emph{et~al.}
\newblock \bibinfo{title}{Designing nanostructures for phonon transport via
  bayesian optimization}.
\newblock \emph{\bibinfo{journal}{Physical Review X}}
  \textbf{\bibinfo{volume}{7}}, \bibinfo{pages}{021024} (\bibinfo{year}{2017}).

\bibitem{mavadia2017prediction}
\bibinfo{author}{Mavadia, S.}, \bibinfo{author}{Frey, V.},
  \bibinfo{author}{Sastrawan, J.}, \bibinfo{author}{Dona, S.} \&
  \bibinfo{author}{Biercuk, M.~J.}
\newblock \bibinfo{title}{Prediction and real-time compensation of qubit
  decoherence via machine learning}.
\newblock \emph{\bibinfo{journal}{Nature communications}}
  \textbf{\bibinfo{volume}{8}}, \bibinfo{pages}{14106} (\bibinfo{year}{2017}).

\bibitem{wigley2016fast}
\bibinfo{author}{Wigley, P.~B.} \emph{et~al.}
\newblock \bibinfo{title}{{Fast machine-learning online optimization of
  ultra-cold-atom experiments}}.
\newblock \emph{\bibinfo{journal}{Scientific reports}}
  \textbf{\bibinfo{volume}{6}} (\bibinfo{year}{2016}).

\bibitem{barber2012bayesian}
\bibinfo{author}{Barber, D.}
\newblock \emph{\bibinfo{title}{Bayesian reasoning and machine learning}}
  (\bibinfo{publisher}{Cambridge University Press}, \bibinfo{year}{2012}).

\bibitem{amstrup1995genetic}
\bibinfo{author}{Amstrup, B.}, \bibinfo{author}{Toth, G.~J.},
  \bibinfo{author}{Szabo, G.}, \bibinfo{author}{Rabitz, H.} \&
  \bibinfo{author}{Loerincz, A.}
\newblock \bibinfo{title}{Genetic algorithm with migration on topology
  conserving maps for optimal control of quantum systems}.
\newblock \emph{\bibinfo{journal}{The Journal of Physical Chemistry}}
  \textbf{\bibinfo{volume}{99}}, \bibinfo{pages}{5206--5213}
  (\bibinfo{year}{1995}).

\bibitem{geisel2013evolutionary}
\bibinfo{author}{Geisel, I.} \emph{et~al.}
\newblock \bibinfo{title}{Evolutionary optimization of an experimental
  apparatus}.
\newblock \emph{\bibinfo{journal}{Applied Physics Letters}}
  \textbf{\bibinfo{volume}{102}}, \bibinfo{pages}{214105}
  (\bibinfo{year}{2013}).

\bibitem{beil2010light}
\bibinfo{author}{Beil, F.}, \bibinfo{author}{Buschbeck, M.},
  \bibinfo{author}{Heinze, G.} \& \bibinfo{author}{Halfmann, T.}
\newblock \bibinfo{title}{Light storage in a doped solid enhanced by
  feedback-controlled pulse shaping}.
\newblock \emph{\bibinfo{journal}{Physical Review A}}
  \textbf{\bibinfo{volume}{81}}, \bibinfo{pages}{053801}
  (\bibinfo{year}{2010}).

\bibitem{tsubouchi2008rovibrational}
\bibinfo{author}{Tsubouchi, M.} \& \bibinfo{author}{Momose, T.}
\newblock \bibinfo{title}{Rovibrational wave-packet manipulation using shaped
  midinfrared femtosecond pulses toward quantum computation: Optimization of
  pulse shape by a genetic algorithm}.
\newblock \emph{\bibinfo{journal}{Physical Review A}}
  \textbf{\bibinfo{volume}{77}}, \bibinfo{pages}{052326}
  (\bibinfo{year}{2008}).

\bibitem{warren1993coherent}
\bibinfo{author}{Warren, W.~S.}, \bibinfo{author}{Rabitz, H.} \&
  \bibinfo{author}{Dahleh, M.}
\newblock \bibinfo{title}{Coherent control of quantum dynamics: the dream is
  alive}.
\newblock \emph{\bibinfo{journal}{Science}} \textbf{\bibinfo{volume}{259}},
  \bibinfo{pages}{1581--1589} (\bibinfo{year}{1993}).

\bibitem{jin2011surrogate}
\bibinfo{author}{Jin, Y.}
\newblock \bibinfo{title}{Surrogate-assisted evolutionary computation: Recent
  advances and future challenges}.
\newblock \emph{\bibinfo{journal}{Swarm and Evolutionary Computation}}
  \textbf{\bibinfo{volume}{1}}, \bibinfo{pages}{61--70} (\bibinfo{year}{2011}).

\bibitem{jin2005comprehensive}
\bibinfo{author}{Jin, Y.}
\newblock \bibinfo{title}{A comprehensive survey of fitness approximation in
  evolutionary computation}.
\newblock \emph{\bibinfo{journal}{Soft computing}}
  \textbf{\bibinfo{volume}{9}}, \bibinfo{pages}{3--12} (\bibinfo{year}{2005}).

\bibitem{won2005framework}
\bibinfo{author}{Won, K.~S.} \& \bibinfo{author}{Ray, T.}
\newblock \bibinfo{title}{A framework for design optimization using
  surrogates}.
\newblock \emph{\bibinfo{journal}{Engineering optimization}}
  \textbf{\bibinfo{volume}{37}}, \bibinfo{pages}{685--703}
  (\bibinfo{year}{2005}).

\bibitem{rayas2004based}
\bibinfo{author}{Rayas-S{\'a}nchez, J.~E.}
\newblock \bibinfo{title}{{EM-based optimization of microwave circuits using
  artificial neural networks: The state-of-the-art}}.
\newblock \emph{\bibinfo{journal}{IEEE transactions on microwave theory and
  techniques}} \textbf{\bibinfo{volume}{52}}, \bibinfo{pages}{420--435}
  (\bibinfo{year}{2004}).

\bibitem{hacioglu2007fast}
\bibinfo{author}{Hacioglu, A.}
\newblock \bibinfo{title}{Fast evolutionary algorithm for airfoil design via
  neural network}.
\newblock \emph{\bibinfo{journal}{AIAA journal}} \textbf{\bibinfo{volume}{45}},
  \bibinfo{pages}{2196--2203} (\bibinfo{year}{2007}).

\bibitem{kaelbling1996reinforcement}
\bibinfo{author}{Kaelbling, L.~P.}, \bibinfo{author}{Littman, M.~L.} \&
  \bibinfo{author}{Moore, A.~W.}
\newblock \bibinfo{title}{{Reinforcement learning: A survey}}.
\newblock \emph{\bibinfo{journal}{Journal of artificial intelligence research}}
  \textbf{\bibinfo{volume}{4}}, \bibinfo{pages}{237–285}
  (\bibinfo{year}{1996}).

\bibitem{thompson1933on}
\bibinfo{author}{Thompson, W.~R.}
\newblock \bibinfo{title}{On the likelihood that one unknown probability
  exceeds another in view of the evidence of two samples}.
\newblock \emph{\bibinfo{journal}{Biometrika}} \textbf{\bibinfo{volume}{25}},
  \bibinfo{pages}{285--294} (\bibinfo{year}{1933}).

\bibitem{hendrycks2016bridging}
\bibinfo{author}{Hendrycks, D.} \& \bibinfo{author}{Gimpel, K.}
\newblock \bibinfo{title}{Bridging nonlinearities and stochastic regularizers
  with gaussian error linear units}.
\newblock \emph{\bibinfo{journal}{arXiv preprint arXiv:1606.08415}}
  (\bibinfo{year}{2016}).

\bibitem{breiman1996bagging}
\bibinfo{author}{Breiman, L.}
\newblock \bibinfo{title}{Bagging predictors}.
\newblock \emph{\bibinfo{journal}{Machine learning}}
  \textbf{\bibinfo{volume}{24}}, \bibinfo{pages}{123--140}
  (\bibinfo{year}{1996}).

\bibitem{Byrd:2006iv}
\bibinfo{author}{Byrd, R.~H.}, \bibinfo{author}{Lu, P.},
  \bibinfo{author}{Nocedal, J.} \& \bibinfo{author}{Zhu, C.}
\newblock \bibinfo{title}{{A Limited Memory Algorithm for Bound Constrained
  Optimization}}.
\newblock \emph{\bibinfo{journal}{SIAM Journal on Scientific Computing}}
  \textbf{\bibinfo{volume}{16}}, \bibinfo{pages}{1190--1208}
  (\bibinfo{year}{2006}).

\bibitem{Sparkes:2013jw}
\bibinfo{author}{Sparkes, B.~M.} \emph{et~al.}
\newblock \bibinfo{title}{{Gradient echo memory in an ultra-high optical depth
  cold atomic ensemble}}.
\newblock \emph{\bibinfo{journal}{New Journal of Physics}}
  \textbf{\bibinfo{volume}{{15}}}, \bibinfo{pages}{085027}
  (\bibinfo{year}{2013}).

\bibitem{cho2016highly}
\bibinfo{author}{Cho, Y.-W.} \emph{et~al.}
\newblock \bibinfo{title}{{Highly efficient optical quantum memory with long
  coherence time in cold atoms}}.
\newblock \emph{\bibinfo{journal}{Optica}} \textbf{\bibinfo{volume}{3}},
  \bibinfo{pages}{100–107} (\bibinfo{year}{2016}).

\bibitem{sparkes2013gradient}
\bibinfo{author}{Sparkes, B.} \emph{et~al.}
\newblock \bibinfo{title}{{Gradient echo memory in an ultra-high optical depth
  cold atomic ensemble}}.
\newblock \emph{\bibinfo{journal}{New Journal of Physics}}
  \textbf{\bibinfo{volume}{15}}, \bibinfo{pages}{085027}
  (\bibinfo{year}{2013}).

\bibitem{ketterle1993high}
\bibinfo{author}{Ketterle, W.}, \bibinfo{author}{Davis, K.~B.},
  \bibinfo{author}{Joffe, M.~A.}, \bibinfo{author}{Martin, A.} \&
  \bibinfo{author}{Pritchard, D.~E.}
\newblock \bibinfo{title}{{High densities of cold atoms in a dark
  spontaneous-force optical trap}}.
\newblock \emph{\bibinfo{journal}{Physical review letters}}
  \textbf{\bibinfo{volume}{70}}, \bibinfo{pages}{2253} (\bibinfo{year}{1993}).

\bibitem{depue1999unity}
\bibinfo{author}{DePue, M.~T.}, \bibinfo{author}{McCormick, C.},
  \bibinfo{author}{Winoto, S.~L.}, \bibinfo{author}{Oliver, S.} \&
  \bibinfo{author}{Weiss, D.~S.}
\newblock \bibinfo{title}{{Unity occupation of sites in a 3D optical lattice}}.
\newblock \emph{\bibinfo{journal}{Physical review letters}}
  \textbf{\bibinfo{volume}{82}}, \bibinfo{pages}{2262} (\bibinfo{year}{1999}).

\bibitem{mloop-package}
\bibinfo{author}{Hush, M.~R.} \& \bibinfo{author}{Slatyer, H.}
\newblock \bibinfo{title}{{M-LOOP}} (\bibinfo{year}{2017}).
\newblock \urlprefix\url{https://github.com/charmasaur/M-LOOP}.

\bibitem{tensorflow2015-whitepaper}
\bibinfo{author}{Abadi, M.} \emph{et~al.}
\newblock \bibinfo{title}{{TensorFlow}: Large-scale machine learning on
  heterogeneous systems} (\bibinfo{year}{2015}).
\newblock \urlprefix\url{https://www.tensorflow.org/}.
\newblock \bibinfo{note}{Software available from tensorflow.org}.

\end{thebibliography}

\normalsize
\paragraph{Methods}

The geometry presented allows for an atomic ensemble 5cm in length containing approximately 10\textsuperscript{10} atoms at a temperature of $\approx 100 \mu$K. Loading and cooling of thermal atoms into the trap is achieved using Doppler cooling with a field 31MHz red detuned from the D2 F=2 $\rightarrow$ F'=3 cycling transition. A field on resonance with the D2 F=1 $\rightarrow$ F'=2 transition provides repump. A compression phase is then applied over 20ms with 1ms updates, which conventionally comprises of detuning the repump field to induce a temporal dark spot while monotonically increasing the magnetic fields to compress the trapped atoms into a smaller volume. However for the ANN optimised solution we switch the values every 1ms following the sequence description presented in Fig \ref{fig:results}c. Following either approach, polarisation gradient cooling is applied for a further 1ms for sub-Doppler cooling by detuning the trapping fields and switching off the magnetic confinement coils. Previous quantum memory schemes use a particular Zeeman coherence \cite{cho2016highly} which we realise by pumping into the m\textsubscript{f} = +1 magnetic sublevel using a field 40MHz red detuned from the D2 F=2 $\rightarrow$ F'=3 during the application of a bias field applied axially along the ensemble. After a 1ms dead time, which allows for the dissipation of eddy currents induced in the surrounding magnetic materials, we probe the OD of the ensemble using an axially propagating field red detuned from the D1 F=1 $\rightarrow$ F'=2 transition. For optimisation purposes a detuning of 90MHz was used to limit the effects of noise and lensing by the atomic ensemble. Our cost function spans a range of 57\% absorption between the maximum achieved absorption at that detuning and a failure to trap atoms.

Optimisation is implemented by a control type feedback loop. The learner has control over the experimental parameters by feeding values to a field programmable gate array (FPGA) which sets the detuning of trapping and repump fields using acousto-optic modulators while the magnetic field strength is set via a voltage controlled current source. Bounds on the parameters are monitored internally by the learner as well as by the control systems which are implemented in both Python and LabVIEW which transfer data via TCP sockets. Feedback is obtained via an acquistion card (NI 5761) which is passed to the learner to train the ANN and generate new experimental parameters.

The machine learning algorithm is built into the Machine-Learning Online Optimisation Package (M-LOOP) \cite{wigley2016fast,mloop-package}
, with the neural networks implemented using Tensorflow \cite{tensorflow2015-whitepaper}.
 Each of the three networks is initialised using He initialisation (an improved version of Xavier initialisation). Training is performed using mini-batch gradient descent with batches of size 16. Training proceeds in iterations, each consisting of 100 epochs (loops over the full data set). At the start of each iteration, a threshold is calculated as 80\% of the current loss. At the end of that iteration, if the new loss is below that threshold then another iteration is performed. Otherwise, training terminates. The networks use $L^2$ regularisation with a coefficient of $10^{-8}$. Before being passed to the networks, all data are normalised using z-scores (based on the initial training data determined during the first $126$ experimental runs). All hyperparameters (network topology, epochs per training iteration, training threshold and regularisation coefficient) were determined by manually tuning the algorithm on a random simulated $10$-dimensional quadratic landscape.
 
Optimisation batches were performed on different days to determine the extent to which experimental drift affected the optimal solution. The duty cycle of the experiment is approximately 2Hz however, acquisition cannot occur at this rate as the experiment takes multiple runs to reach equilibrium. A single optimisation would generally take approximately 2hrs at which point further acquisition was limited by experimental drift. Many solutions were identified with slightly different parameter profiles. Each was verified to be a local minima by manual tuning around the optimal solution which invariably resulted in a reduction in OD. We also note that each solution retains its relative effectiveness compared with other solutions arrived at by a given optimisation, with the presented solution corresponding to the highest OD during all optimisation batches.
\end{document}